\documentclass[12pt]{article}

\usepackage{amssymb,amsfonts,amsmath}
\usepackage{longtable}
\usepackage{graphicx}
\usepackage{epsfig}
\usepackage{subfig}
\usepackage{url}

\newtheorem{theorem}{Theorem}[section]

\newtheorem{remark}{Remark}[section]

\begin{document}

\title{Testing for replicability in a follow-up study when the primary study hypotheses are two-sided}
\maketitle

\begin{center}
Ruth Heller \\
\emph{Department of Statistics and Operations Research, Tel-Aviv
university, Tel-Aviv, Israel. E-mail: ruheller@post.tau.ac.il}\\
Marina Bogomolov \\
\emph{Faculty of Industrial  Engineering and Management, Technion --
Israel Institute of Technology, Haifa, Israel. E-mail:
marinabo@tx.technion.ac.il }\\
Yoav Benjamini \\
\emph{Department of Statistics and Operations Research, Tel-Aviv
university, Tel-Aviv, Israel. E-mail: ybenja@post.tau.ac.il}\\
Tamar Sofer\\
\emph{Department of Biostatistics, University of Washington, Seattle, USA. E-mail:  tsofer@uw.edu}
\end{center}

\begin{abstract}
When testing for replication of results from a primary study with two-sided hypotheses in a follow-up study, we are usually interested in discovering the features with discoveries in the same direction in the two studies. The direction of testing in the follow-up study for each feature can therefore be decided by the primary study.  
We prove that in this case the methods suggested in \cite{Heller14b} for  control over false replicability claims are valid. Specifically, we prove that if we input  into the procedures in \cite{Heller14b} the one-sided $p$-values in the directions favoured by the primary study, then we achieve directional control over the desired error measure (family-wise error rate or false discovery rate). 
\end{abstract}

\vspace{0.5cm}

\section{Introduction}

In this note we are concerned with the setting where two studies (a primary study and a follow-up study) are available that examine the same problem, and the aim is to discover the features with true findings with directional consistency from the primary to the follow-up study. So features with true left-sided alternatives in both studies are of interest, and features with true right-sided alternatives in both studies are of interest, but a feature with a true left-sided alternative in one of the studies and a true right-sided alternative in the other study is not. 

In  \cite{Heller14b} we suggested procedures for declaring that findings from a primary study have been replicated in a follow-up study, where a feature has a  false replicability claim if the null hypothesis is true in at least one of the studies.
Our proposal assigned an $r$-value to each finding. The false discovery rate (FDR)  $r$-value for feature $i$ is the lowest FDR  level  at which we can say that the finding is among the replicated ones. We showed that the the procedure that declares findings with FDR $r$-values below $q$ as replicated controls the FDR  on replicability claims. For family-wise error-rate (FWER) control on replicability claims, we suggested using FWER $r$-values with a similar property. 
In this note, we extend the results in \cite{Heller14b} for the setting that the direction of
the effects is unknown in advance, and the direction of
 testing is determined by the data of the primary study. We suggest using for each feature the minimum one-sided primary study $p$-value, i.e., we consider for replicability the direction the data favours, and in the follow-up the one-sided $p$-value in the favoured direction determined by the primary study. For example, if in a primary genome-wide association study (GWAS) the direction of association is unknown in advance, the primary study serves to guide two important design decisions for follow-up: first, which hypotheses will be followed-up, and second, the direction of testing in the follow-up study. For replicability analysis, we will need the one-sided $p$-values in the primary and follow-up studies, in the direction of association determined by the primary study.
Although  we decide for each feature the  direction of testing based on its test statistic in the primary study, if we compute the $r$-values as in \cite{Heller14b}, then the procedures that declare as replicated all features with all $r$-values below the nominal level control the directional FDR/FWER.  This is remarkable, since we are used to paying a factor of two for a
single study when the direction of testing is unknown:  we multiply the (minimal) one-sided $p$-value for two-sided hypothesis testing. For replicability analysis, there is no cost of not knowing before looking at the results from the primary study the direction of testing for replicability. The reason is that these procedures already have a cost for the fact that we select the promising hypotheses from the primary to the follow-up, and this cost actually covers also the selection of direction of interest.
The $r$-values can be computed using our web application \url{http://www.math.tau.ac.il/$\sim$ruheller/App.html}. An R script is also available in RunMyCode,\url{http://www.runmycode.org/companion/view/542}.


\section{Notation}\label{notation}
Here we give the formal framework for replicability analysis for two-sided hypotheses, including directional errors.
Consider a family of $m\geq 1$ features examined in the primary study.
Feature $j\in \{1,\ldots,m\}$ in study $i\in \{1,2\}$ has either a true or a false null hypothesis. If the null hypothesis is false, the true left-sided or right-sided alternative is true. We define $H_{ij}$ as follows:
\begin{equation*} H_{ij}
= \left\{
\begin{array}{rl}
1 & \text{if the right-sided alternative is true for feature $j$ in study $i$},\\
0 & \text{if the null hypothesis is true for feature $j$  in study $i$},\\
-1 & \text{if the left-sided alternative is true for feature $j$ in study $i$}.\\
\end{array} \right.
\end{equation*}

Let $\mathcal{H} = \{\vec{h} = (h_1,h_2): h_i \in \{-1,0,1 \}\}$ be the set of $9$ possible configurations of the vector $\vec H_j = (H_{1j}, H_{2j})$ for two-sided alternatives. (If interest lies only in detecting left or right sided alternatives, only the relevant subset of $4$ possible configurations are considered.)
The set of $m$ features can be divided into $9$ (unknown) subsets defined by $\mathcal{H}: \{j: \vec{H}_j=\vec{h}\}, \vec{h}\in\mathcal{H}$.
Each feature $j$ is in exactly one of the 9 subsets. 
The subsets of features whose effect is not replicated in the same direction in the two studies are $\{j:\vec{H_j}=\vec{h}\}$, for $\vec{h}\in \mathcal{H}_0=\{(-1,1),
(-1,0), (1,0), (0,0), (0,1), (0.-1), (1,-1)\}.$ The goal in inference is to discover as many features as possible with $\vec H_j \notin \mathcal{H}^0 $, i.e., $\vec H_j \in \{(1,1), (-1,-1) \}$.

Suppose  $R$ replicability claims are made by an analysis. Let $R_j^L$ and $R_j^R$ be the indicators of whether a replicability claim was made for feature $j$ in the left and right direction, respectively. The number of  replicability claims that are true (i.e., that are with true left-sided alternatives in both studies or true right-sided alternatives in both studies) is $$S = \sum_{\{j: \vec H_j = (1,1)\}}R_j^R+\sum_{\{j: \vec H_j = (-1,-1)\}}R_j^L,$$  and $R-S$ is the number of  replicability claims that are false (i.e., that are with true left-sided alternatives in at most one study and true right-sided alternatives in at most one study).

 The directional FWER criterion is the probability of at least one false directional replicability claim,
 $$FWER = \textmd{Pr}(R-S>0).$$
The directional FDR for replicability analysis is the expected proportion of false
directional replicability claims among all those called replicated:
$$FDR = E\left(\frac{R-S}{\max (R,1)}\right).$$

For feature $j$ tested in the follow-up study, the left- and right- sided $p$-values for study $i\in \{1,2\}$ are denoted by $p^L_{ij}$ and $p^R_{ij}$, respectively. For continuous test statistics, $p^R_{ij} = 1-p^L_{ij}$. For replicability analysis, we will need the one-sided $p$-values only in the direction favoured by the primary study, i.e. the pair $(p'_{1j}, p'_{2j})$ defined as follows:
\begin{equation*}
p'_{1j} = \left\{
\begin{array}{rl}
p_{1j}^L & \text{if }  p_{1j}^L< p_{1j}^R,\\
p_{1j}^R & \text{if }  p_{1j}^L> p_{1j}^R.
\end{array} \right.
\quad
p'_{2j} = \left\{
\begin{array}{rl}
p_{2j}^L & \text{if }  p_{1j}^L< p_{1j}^R,\\
p_{2j}^R & \text{if }  p_{1j}^L> p_{1j}^R.
\end{array} \right.
\end{equation*}

\begin{remark}
We implicitly assume that $p_{1j}'\leq 0.5$. This assumption may be violated if the test statistic is discrete. Since features with $p_{1j}'> 0.5$ are obviously not interesting features in the primary study, we suggest including in our selection rule (of which features to follow-up on) the condition that a feature is selected only if $p_{1j}'\leq 0.5$. 
\end{remark}

\section{FDR Replicability from follow-up for two-sided hypotheses}
As in \cite{Heller14b}, we let $f_{00}$ denote the  fraction of features, out of the $m$ features examined in the primary study,  that are null in both studies. We cannot estimate $f_{00}$ from the data, since only a handful
 of promising features  are followed up in practice. However, $f_{00}$ is typically closer to one than to
zero, and  we can give a conservative guess for a lower bound on $f_{00}$, call it $l_{00}$.  For example, in typical GWAS on the whole genome, $l_{00}=0.8$ is conservative. We can exploit the fact that $l_{00}>0$ to gain power.

\subsection{Computation of $r$-values for FDR-replicability for two-sided hypotheses}\label{subsec-compFDRrvalues}
For completeness, we present the procedure for establishing replicability from follow-up, which is identical to the procedure in \cite{Heller14b}. The only difference is that the one-sided $p$-values to input into the procedure are the ones favoured by the primary study. 

\begin{enumerate}
\item Data input:
\begin{enumerate}
\item $m$, the number of features examined in the primary study.
\item $\mathcal R_1$, the set of features selected for follow-up based on primary study results. Let $R_1 = |\mathcal R_1|$ be their number.
\item $\{(p'_{1j},p'_{2j}): j\in\mathcal R_1\}$, where  $p'_{1j}$ and $p'_{2j}$ are, respectively, the primary and follow-up study one-sided $p$-values for feature  $j\in\mathcal R_1$ in the direction favoured by the primary study.
\end{enumerate}
\item Parameters input:
\begin{enumerate}
\item $l_{00}\in [0,1)$, the lower bound on $f_{00}$ (see above), default value for whole genome GWAS is $l_{00}=0.8$.
\item $c_2\in (0,1)$, the emphasis given to the follow-up study (see Section Variations in \cite{Heller14b}), default value is $c_2 =0.5$.
\end{enumerate}
\item Definition of the functions $f_i(x), i \in \mathcal R_1, x\in (0,1)$:
\begin{enumerate}
\item Compute
$c_1(x)  = \frac{1-c_2}{1-l_{00}(1-c_2x)}$.
\item For every feature $j\in \mathcal R_1 $ compute the following $e$-values:\\
$$e_j(x)=\max\left(\frac{1}{c_1(x)}p'_{1j},\,\frac{R_1}{m c_2}p'_{2j}\right), \quad j
\in \mathcal{R}_1.$$
\item Let $f_i(x) = \min_{\{j: e_j(x)\geq e_i(x), j \in \mathcal R_1\}}\frac{e_j(x)
m }{rank[e_j(x)]}$, where $rank[e_j(x)]$ is the rank of the $e$-value for feature $j\in \mathcal R_1$ (with maximum rank for ties).
\end{enumerate}
\item The FDR $r$-value for feature $i\in \mathcal R_1$ is the solution to $f_i(r_i) = r_i$ if a solution exists in $(0,1)$,  and 1 otherwise. The solution is unique, see Supplementary Information (SI) Lemma S1.1 in \cite{Heller14b} for a proof.
\end{enumerate}

\subsection{The level $q$ directional FDR replicability procedure}
\begin{enumerate}
\item Compute the $r$-values as detailed in Section \ref{subsec-compFDRrvalues}.
\item  The replicability claims at a prefixed level $q$, say $q=0.05$, are all features with $r$-values at most $0.05$. Denote this  set of features by $\mathcal R_2$.
\item If feature $j\in \mathcal R_2 $ has $p'_{1j}=p^L_{1j}$, then  declare the feature as having a replicated true left-sided alternative (i.e., a true  effect/signal/association in the left direction) ; If feature $j\in \mathcal R_2$ has $p'_{1j}=p^R_{1j}$, then  declare the feature as having a replicated true right-sided alternative  (i.e., a true effect/signal/association in the right direction).
\end{enumerate}

The directional FDR for replicability analysis is then controlled at level 0.05, as we show in Theorem \ref{theorem-upperbound1}, as long as the selection rule is stable.

{\noindent \bf Definition \cite{Heller14b}. }{\sl  A stable selection rule
satisfies the following condition: for any $j\in \mathcal{R}_1$, changing $p^L_{1j}$ so
that $j$ is still selected while all other primary study $p$-values are held fixed, will not change the set $\mathcal R_1$.}

Stable selection rules include
 selecting the hypotheses with two-sided primary $p$-values below a certain cut-off, or by a
 non-adaptive multiple testing procedure on the primary study two-sided $p$-values such as the BH procedure
 for FDR control or the Bonferroni procedure for FWER control, or selecting the $k$ hypotheses with the smallest two-sided
 $p$-values, where $k$ is fixed in advance.

\begin{theorem}\label{theorem-upperbound1}
A procedure that declares findings with $FDR$ $r$-values at most $q$ as replicated controls the directional FDR for replicability analysis at level  at most $q$ if the following conditions are satisfied: 
the rule by which the set $\mathcal{R}_1$ is selected is a stable selection rule; $l_{00}\leq f_{00}$; the $p$-values
within the follow-up study are jointly independent or are positive regression dependent on the subset of $p$-values corresponding to true null hypotheses (property
PRDS\footnote{Property PRDS was introduced in \cite{Benjamini01}. For example, the PRDS property is satisfied if the test statistics are Gaussian, non-negatively correlated, and the tested hypotheses are one-sided.});  for features with $\vec H_j \notin \{(1,1), (-1,-1)\}$ the follow-up study $p$-values are independent of the primary study $p$-values; and in addition one of items 1-3 below is satisfied. 
\begin{enumerate}
\item The $p$-values within the primary study are
independent.
\item Arbitrary dependence among the $p$-values within the primary
study, when in Step 3 in Section \ref{subsec-compFDRrvalues} $m$ is replaced by
$m^*=m\sum_{i=1}^m1/i.$
\item Arbitrary dependence among the $p$-values within the primary study, and the selection rule is such that the primary study $p$-values of the features that are selected for follow-up
are at most a fixed threshold $t\in (0,1)$, when $c_1$  computed in Step 3(a) is replaced by
$$\tilde{c}_1(x)=\max\{a:\,a(1+\sum_{i=1}^{\lceil tm/(ax)-1\rceil}1/i)=c_1(x)\}.$$
 Steps 3(b) and 3(c) remain unchanged. In step 4, the FDR $r$-value for feature $i\in \mathcal R_1$  is $r_i= \min\{x: f_i(x)\leq x\}$ if a solution exists in $(0,1)$, and one otherwise.
\end{enumerate}
\end{theorem}
See Appendix \ref{app-FDRproof} for a proof. The implication of item
3 is that for FDR-replicability at level $q$, if $t\leq c_1(q)q/m$,
no modification is required, so the procedure that declares as
replicated all features with $r$-values at most $q$ controls the FDR
at level $q$ on replicability claims for any type of dependency in
the primary study. Note that the modification in item 3 will lead to
more discoveries than the modification in item 2 only if
$t<\frac{c_1(q)q}{1+\sum_{i=1}^{m-1}1/i}$.

 We conjecture from empirical investigations that even if the primary study $p$-values are not independent, the conservative modifications of the $r$-value computation in items 2-3 are unnecessary for FDR control in replicability analysis of GWAS studies, see \cite{Heller14b} for details.


\section{FWER Replicability from follow-up for two-sided hypotheses}
\subsection{Computation of $r$-values for FWER-replicability for two-sided hypotheses}\label{subsec-FWER}
 The directional FWER
criterion, $$FWER = \textmd{Pr}(R-S>0),$$ is more
stringent than the directional FDR, yet it may sometimes be desired. 
The directional FWER $r$-value is the lowest directional FWER level at which we can say that the finding has been  replicated. The $r$-value can be compared to any desired level of directional FWER.
For feature $j\in \mathcal{R}_1$,
$$f^{Bonf}_j(x) = m \cdot e_j(x),$$
where $e_j(x)$ is the $e$-value defined in Step 3(b) of Section \ref{subsec-compFDRrvalues}.
The Bonferroni $r$-value for feature $j$ is the solution to $f^{Bonf}_j(r_j) = r_j$ if a solution exists in $[0,1)$, and one otherwise. It can be shown that the solution is unique 
similarly to the case with FDR $r$-values.

\subsection{The level $\alpha$ directional FWER-replicability procedure}\label{subsec-directionalFWER}
\begin{enumerate}
\item Compute the $r$-values as detailed in Section \ref{subsec-FWER}.
\item  The replicability claims at a prefixed level $\alpha$, say $\alpha=0.05$, are all features with $r$-values at most $\alpha$. Denote this  set of features by $\mathcal R_2$.
\item If feature $j\in \mathcal R_2 $ has $p_{1j}^L<p_{1j}^R$, then  declare the feature as having a replicated true left-sided alternative; If feature $j\in \mathcal R_2$ has $p_{1j}^R<p_{1j}^L$, then  declare the feature as having a replicated true right-sided alternative.
\end{enumerate}
The directional FWER for replicability analysis is then controlled at level 0.05, as stated in the following theorem.

\begin{theorem}\label{thm_FWER}
The procedure above controls the directional FWER for replicability
analysis at level $\alpha$ if $l_{00}\leq f_{00}$, and if for features with $\vec H_j \notin \{(1,1), (-1,-1)\}$ the follow-up study  $p$-values are independent of the primary study $p$-values.
\end{theorem}

See Appendix \ref{app-FWERproof} for the proof.


\

\appendix

\section{Proof of Theorem \ref{theorem-upperbound1}}\label{app-FDRproof}
We first show that the following procedure is identical to that of
declaring the set of findings with FDR-replicability $r$-values  at
most $q$ as replicated. First, compute the number of replicability
claims at level $q$ as follows:
 $$R_2\triangleq\max\left\{r:
\sum_{j\in\mathcal{R}_1}\textbf{I}\left[(p'_{1j},
p'_{2j})\leq\left(\frac{r}{m}c_1(q)q, \frac{r}{R_1}c_2
q\right)\right] = r\right\}.$$ Next, declare as replicated findings
the set
$$\mathcal R_2= \left\{j: (p'_{1j}, p'_{2j})\leq\left(\frac{R_2}{m}c_1(q)q,
\frac{R_2}{R_1}c_2q\right), j \in \mathcal R_1\right\}.$$ It was
shown in Lemma S1.1 in \cite{Heller14b} that when the hypotheses are
one-sided, this procedure is identical to declaring the set of
findings with FDR-replicability $r$-values at most $q$ as
replicated, when the one-sided $p$-values replace $(p'_{1j},
p'_{2j})$ both in the above procedure and in the computation of
FDR-replicability $r$-values. It is straightforward to see that the
proof of Lemma S1.1 in \cite{Heller14b} remains unchanged when the
one-sided $p$-values are replaced by $(p'_{1j}, p'_{2j}),$ therefore
the above procedure is identical  to that of declaring the set of
findings with FDR-replicability $r$-values  at most $q$ as
replicated for two-sided hypotheses.

We shall prove that under the conditions of items 1-3 of Theorem \ref{theorem-upperbound1} the above procedure controls the FDR for replicability analysis  at a level
which is smaller or equal to
\begin{eqnarray}
&&c_1(q)c_2q^2(|j: \vec H_j \in \{(-1,0),(1,0),(0,0) \}|)/m+\nonumber\\
&&c_1(q)q|j: \vec H_j \in \{(0,1),(0,-1),(-1,-1),(1,1),(-1,1),(1,-1)\}|/m+\nonumber \\
&&c_2q E[|\mathcal R_1 \cap \{j:\vec H_j \in
\{(-1,0),(1,0),(-1,1),(1,-1),(0,1),(0,-1),(0,0) \}| /|\mathcal R_1|]
,\nonumber \\\label{upper-FWER}
\end{eqnarray}
where the cardinalities are over the sets containing all $m$
features, i.e. $j=1,\ldots,m$. Note that this expression is at most
$q$ if $l_{00} \leq f_{00}$. To see this, note that
$$|j: \vec H_j \in \{(-1,0),(1,0),(0,0) \}|/m = f_{\cdot 0},$$
and
$$|j: \vec H_j \in \{(0,1),(0,-1),(-1,-1),(1,1),(-1,1),(1,-1) \}|/m = 1-f_{\cdot 0}.$$
Moreover, $$E[|\mathcal R_1 \cap \{j:\vec H_j \in
\{(-1,0),(1,0),(-1,1),(1,-1), (0,1), (0,-1),(0,0) \}| /|\mathcal
R_1|]\leq 1.$$ Therefore, expression (\ref{upper-FWER}) is at most
\begin{eqnarray}
&&c_1(q)c_2q^2f_{\cdot 0}+c_1(q)q(1-f_{\cdot 0})+c_2q\nonumber\\
&&=c_1(q)q-f_{\cdot 0}c_1(q)q(1-c_2q) + c_2q \nonumber\\
&&\leq c_1(q)q-l_{00}c_1(q)q(1-c_2q) + c_2q \nonumber\\
&& = c_1(q)q[1-l_{00}(1-c_2q)]+ c_2q \nonumber \\
&& = (1-c_2)q + c_2q = q. \nonumber
\end{eqnarray}

We shall now prove that the expression in (\ref{upper-FWER}) is an
upper bound for the directional FDR for replicability analysis,
which is
\begin{align}
&E\left(\frac{R-S}{\max (R,1)}\right)=\notag\\&\sum_{\{j:\vec H_j
\in \{(0,-1),(0,1),(0,0),(1,0),(-1,0),(1,-1),(-1,1)\}
\}}E\left(\frac{R_j^L+R_j^R}{\max(R,
1)}\right)+\notag\\&\sum_{\{j:\vec H_j=(1,1)
\}}E\left(\frac{R_j^L}{\max(R, 1)}\right)+\sum_{\{j:\vec H_j=(-1,-1)
\}}E\left(\frac{R_j^R}{\max(R,1)}\right).\label{fdr}
\end{align}
For each $j\in\{1,\ldots,m\},$ we define $C_r^{(j)}$ as the event in
which if $j$ is declared replicated, $r$ hypotheses are declared
replicated including $j$, which amounts to the definition given in
the proof of Theorem 1 in Supplementary Material of
\cite{Heller14b}, where the one-sided $p$-values $(p_{1j}, p_{2j})$
are replaced by $(p'_{1j}, p'_{2j}).$ Note that for any given realization  of $|\mathcal{R}_1|$ and value of $r$ such that $r>|\mathcal{R}_1|$, $C_r^{(j)} = \emptyset.$

From the equivalent procedure above the following equality
follows.
\begin{align}
E\left(\frac{R_j^L}{\max(R,1)}\right) &= \sum_{r=1}^m \frac 1r
\textmd{Pr}\left(j \in \mathcal{R}_1, P_{1j}^L\leq
\min\left(\frac{rc_1(q)q}{m}, 0.5\right), P_{2j}^L\leq
\frac{rc_2q}{\max(|\mathcal{R}_1|, 1)},
C_r^{(j)}\right)\notag\\&\leq \sum_{r=1}^m \frac 1r
\textmd{Pr}\left( P_{1j}^L\leq \frac{rc_1(q)q}{m}, P_{2j}^L\leq
c_2q, C_r^{(j)}\right),\label{indepc1}
\end{align}
where the equality follows from the fact that a replicability claim
is made in the left direction only if $P_{1j}^L\leq P_{1j}^R$, i.e.
only if $P_{1j}^L<0.5.$ 
Similarly,
\begin{align}
E\left(\frac{R_j^R}{\max(R,1)}\right) 
\leq \sum_{r=1}^m \frac 1r
\textmd{Pr}\left( P_{1j}^R\leq \frac{rc_1(q)q}{m}, P_{2j}^R\leq
c_2q, C_r^{(j)}\right).\label{indepc2}
\end{align}
Using inequalities (\ref{indepc1})  and (\ref{indepc2}), 
and the
facts that $P_{1j}^L$ and $P_{1j}^R$ are uniform for
$j\in\{j:H_{1j}=0\}$ and are stochastically larger than uniform for
$j\in\{j:H_{1j}=1\}$ and $j\in\{j:H_{1j}=-1\}$ respectively, 
we obtain the following inequalities:
\begin{equation*}
E\left(\frac{R_j^L}{\max(R,1)}\right) \leq  \left\{
\begin{array}{ll}
c_1(q)q/m & \text{if }  \vec H_j\in\{(0,-1), (1,-1), (1,1)\},\\
c_2q E[I(j\in \mathcal R_1)/|\mathcal R_1|] & \text{if } \vec H_j \in \{(-1,0),(0,1), (-1,1) \},\\
 c_1(q)c_2q^2/m & \text{if } \vec H_j \in \{(0,0),(1,0) \},
\end{array} \right.
\end{equation*}
\begin{equation*}
E\left(\frac{R_j^R}{\max(R,1)}\right) \leq  \left\{
\begin{array}{ll}
c_1(q)q/m & \text{if }  \vec H_j\in\{(0,1), (-1,1), (-1,-1)\},\\
c_2q E[I(j\in \mathcal R_1)/|\mathcal R_1|] & \text{if } \vec H_j \in \{(1,0),(0,-1), (1,-1), (0,0)\},\\
 c_1(q)c_2q^2/m & \text{if } \vec H_j \in \{(-1,0) \}.
\end{array} \right.
\end{equation*}
These upper bounds for items 1-3 of Theorem
\ref{theorem-upperbound1} follow from similar derivations to these
given in the proof of items (\textit{i})-(\textit{iii}) of Theorem 1
in \cite{Heller14b}, respectively. Specifically, for each of the
items, the upper bounds $c_1(q)q/m,$ $c_2q E[I(j\in \mathcal
R_1)/|\mathcal R_1|]$ and $c_1(q)c_2q^2/m$ are derived similarly to
inequalities [S3], [S4], and [S5] in the proof of Theorem 1 in
\cite{Heller14b}, respectively. Thus we obtain
\begin{equation*}
E\left(\frac{R_j^R + R_j^L}{\max(R, 1)}\right) \leq  \left\{
\begin{array}{rl}
c_2q E[I(j\in \mathcal R_1)/|\mathcal R_1|]+c_1(q)c_2q^2/m  & \text{if }  \vec H_j=(0,0),\\
c_2q E[I(j\in \mathcal R_1)/|\mathcal R_1|]+c_1(q)c_2q^2/m & \text{if } \vec H_j \in \{(1,0),(-1,0) \},\\
c_1(q)q/m+c_2q E[I(j\in \mathcal R_1)/|\mathcal R_1|] & \text{if } \vec H_j \in \{(0,1),(0,-1) \},\\
c_2q E[I(j\in \mathcal R_1)/|\mathcal R_1|]+c_1(q)q/m & \text{if }
\vec H_j \in \{(1,-1),(-1,1) \},
\end{array} \right.
\end{equation*}
and for the directional error terms:
\begin{eqnarray}
&& E\left(\frac{R_j^L}{\max(R, 1)}\right)\leq \frac{c_1(q)q}{m}, \quad \textrm{for $j$ with} \quad  \vec H_j=(1,1)  \nonumber \\
&& E\left(\frac{R_j^R}{\max(R, 1)}\right)\leq \frac{c_1(q)q}{m},
\quad \textrm{for $j$ with} \quad \vec H_j=(-1,-1).  \nonumber
\end{eqnarray}
The result follows from using expression (\ref{fdr}) for the
directional FDR for replicability analysis, and summing up over the
above upper bounds.

\section{Proof of Theorem \ref{thm_FWER}}\label{app-FWERproof}
It is easy to show that the procedure in Section
\ref{subsec-directionalFWER} is unchanged if we replace Step 2 by
the following: the replicability claims are all  features with
$f_j^{Bonf}(\alpha)\leq \alpha$. The equivalence follows from the
facts that $f_j^{Bonf}(x)$ is a continuous function of $x$ and
$f_j^{Bonf}(x)/x$ is strictly monotone decreasing (this result
follows from the proof of Lemma S1.1 
in the SI of \cite{Heller14b} and it is straightforward to show
that it continues to hold in the directional replicability
analysis).


We shall now prove that the expression in (\ref{upper-FWER}) with $q$ replaced by $\alpha$ is an
upper bound for the directional FWER for replicability analysis,
which is $\textmd{Pr}(R-S>0).$ It was shown in the proof of Theorem \ref{theorem-upperbound1} that this expression is at most $\alpha$ if $l_{00}\leq f_{00}.$ Note that
\begin{eqnarray}
&&\textmd{Pr}(R-S>0)\leq E(R-S)\leq \sum_{\{j:\vec H_j=(1,1) \}}E(R_j^L)+\sum_{\{j:\vec H_j=(-1,-1) \}}E(R_j^R) \nonumber \\
&& +\sum_{\{j:\vec H_j \in
\{(0,-1),(0,1),(0,0),(1,0),(-1,0),(1,-1),(-1,1)\} \}} E(R_j^R +
R_j^L) \nonumber
\end{eqnarray}

We consider the equivalent procedure that  replaces Step 2 by
counting as replicability claims all  features with
$f_j^{Bonf}(\alpha)\leq \alpha$ (as discussed above). The
directional error terms in the first two sums contribute the
following:
\begin{eqnarray}
&& E\left(R_j^L\right)\leq \frac{c_1(\alpha)\alpha}{m}, \quad \textrm{for $j$ with} \quad  \vec H_j=(1,1)  \nonumber \\
&& E\left(R_j^R\right)\leq \frac{c_1(\alpha)\alpha}{m}, \quad
\textrm{for $j$ with} \quad  \vec H_j=(-1,-1)  \nonumber
\end{eqnarray}
To see how these upper bounds were derived, we consider only the
first (since the second is derived similarly). For $j$ with
$\vec H_j=(1,1)$
\begin{eqnarray}
&& E\left(R_j^L\right)\leq \textmd{Pr}(P_{1j}^L\leq \min(c_1(\alpha)\alpha/m,0.5), P_{2j}^L\leq c_2\alpha/R_1) \nonumber \\
&& \leq \textmd{Pr}(P_{1j}^L\leq c_1(\alpha)\alpha/m) \leq c_1\alpha/m,
\nonumber
\end{eqnarray}
where the first inequality follows from the fact that a
replicability claim is made in the left direction only if
$P_{1j}^L\leq P_{1j}^R$, i.e., only if $P_{1j}^L<0.5$,  and the last
inequality follows that the fact that for $H_{1j}=1$, $P_{1j}^L$ is
stochastically larger than uniform.

All remaining errors are false replicability claims, not only
directional errors. Clearly,
$$E(R_j^R + R_j^L) = \textmd{Pr}(\min(P_{1j}^L, P_{1j}^R)\leq c_1(\alpha)\alpha/m, P'_{2j}\leq c_2 \alpha/|\mathcal R_1|, j\in \mathcal{R}_1).$$
It is simple to show (using similar derivations to these in the proof of Theorem S6.1 in the SI of
\cite{Heller14b}) that the right hand side is at most the following
upper bounds:

\begin{equation*}
E(R_j^R + R_j^L) \leq  \left\{
\begin{array}{rl}
c_2\alpha E[I(j\in \mathcal R_1)/|\mathcal R_1|]+c_1(\alpha)\alpha/m\times c_2 \alpha  & \text{if }  \vec H_j=(0,0),\\
c_2\alpha E[I(j\in \mathcal R_1)/|\mathcal R_1|]+c_1(\alpha)\alpha/m\times c_2 \alpha & \text{if } \vec H_j \in \{(1,0),(-1,0) \},\\
c_1(\alpha)\alpha/m+c_2\alpha E[I(j\in \mathcal R_1)/|\mathcal R_1|] & \text{if } \vec H_j \in \{(0,1),(0,-1) \},\\
c_2\alpha E[I(j\in \mathcal R_1)/|\mathcal R_1|]+c_1(\alpha)\alpha/m
& \text{if } \vec H_j \in \{(1,-1),(-1,1) \}.
\end{array} \right.
\end{equation*}

The result follows from summing over these upper bounds.


\begin{thebibliography}{}
\bibitem{Heller14b}
Heller, R. and Bogomolov, M. and Benjamini, Y. (2014).
\newblock Deciding whether follow-up studies have replicated findings in a preliminary large-scale “omics’ study,
\newblock  {\em Proceedings of the National Academy of Sciences}.

\bibitem{yoav1}
Benjamini, Y. and Hochberg, Y. (1995).
\newblock Controlling the false discovery rate - a practical and powerful
  approach to multiple testing.
\newblock {\em J. Roy. Stat. Soc. B Met.}, 57 (1):289--300.

\bibitem{Benjamini01}
Benjamini, Y. and Yekutieli, D. (2001).
\newblock The control of the false discovery rate in multiple testing under dependency. 
\newblock {\em The Annals of Statistics}, 29 (4):1165--1188.


\bibitem{Reiner03}
Reiner, A. and Yekutieli, D. and Benjamini, Y.  (2003).
\newblock Identifying differentially expressed genes using false discovery rate controlling procedures.
\newblock {\em Bioinformatics}, 19 (3): 368--375.





\bibitem{Bogomolov12}
Bogomolov, M. and Heller, R. (2013).
\newblock Discovering findings that replicate from a primary study of high
  dimension to a follow-up study.
\newblock {\em Journal of the American Statistical Association}, DOI: 10.1080/01621459.2013.829002.


\bibitem{Heller13}
Heller, R. and Yekutieli, D. (2013).
\newblock Replicability analysis for genome-wide
association studies.
\newblock {\em The Annals of Applied Statistics}, accepted.



\end{thebibliography}
\end{document}